\documentclass[conference]{IEEEtran}
\usepackage[  top=0.75in,  bottom=1.08in,  left=0.625in,  right=0.625in]{geometry}
\usepackage{cite}
\usepackage{amsmath,amssymb,amsfonts}
\usepackage{graphicx}
\usepackage{textcomp}
\usepackage{xcolor}
\usepackage{comment}
\usepackage{algorithm}
\usepackage{algpseudocode}  

\graphicspath{ {./Images/} }
\begin{document}

\title{Full-Duplex Beamforming Optimization for Near-Field ISAC}

\author{
\begin{tabular}{ccc}
Ahsan Nazar & Zhambyl Shaikhanov & Sennur Ulukus \\
\multicolumn{3}{c}{Department of Electrical and Computer Engineering} \\
\multicolumn{3}{c}{University of Maryland, College Park, MD 20742, USA} \\
\textit{anazar@umd.edu} & \textit{zhambyl@umd.edu} & \textit{ulukus@umd.edu} \\
\end{tabular}
}

\maketitle

\begin{abstract}
Integrated Sensing and Communications (ISAC) is a promising technology for future wireless networks, enabling simultaneous communication and sensing using shared resources. This paper investigates the performance of full-duplex (FD) communication in near-field ISAC systems, where spherical-wave propagation introduces unique beam-focusing capabilities. We propose a joint optimization framework for transmit and receive beamforming at the base station to minimize transmit power while satisfying rate constraints for multi-user downlink transmission, multi-user uplink reception, and multi-target sensing. Our approach employs alternating optimization combined with semi-definite relaxation and Rayleigh quotient techniques to address the non-convexity of the problem. Simulation results demonstrate that FD-enabled near-field ISAC achieves superior power efficiency compared to half-duplex and far-field benchmarks, effectively detecting targets at identical angles while meeting communication requirements.
\end{abstract}


\section{Introduction}
Integrated Sensing and Communications (ISAC) is a key technique for future 6G wireless networks, combining communication and sensing to improve spectral efficiency and reduce costs by sharing hardware resources \cite{liuIntegratedSensingCommunications2022}. ISAC supports diverse applications in commercial, military, and public safety sectors, such as the Internet of Vehicles (IoV), to both detect and communicate among vehicles and roadside stations.
To meet demands for reliable sensing and efficient communication, ISAC systems will adopt extremely large-scale antenna arrays (ELAAs) and millimeter-Wave (mmWave) or higher frequencies \cite{bjornsonMassiveMIMOReality2019}, necessitating a shift from far-field (planar-wave) to near-field (spherical-wave) channel models \cite{7942128,liuNearFieldCommunicationsWhat2023}.


The spherical-wave propagation introduces a new \textit{distance} dimension in addition to the traditional angular domain, resulting in a beam-focus effect \cite{quNearFieldIntegratedSensing2023}, allowing transmissions to be directed to specific locations rather than just angles. This enables simultaneous estimation of distance and angle in sensing applications and mitigates interference encountered in communications \cite{zhangBeamFocusingNearField2022}. Specifically, the beam-focus effect allows the detection of targets in the same direction, which is impossible in the far-field scenario. 

Most prior ISAC studies focus on far-field models, which are insufficient for near-field scenarios due to significant phase discrepancies and performance loss. This limitation necessitates a fundamental redesign of ISAC systems for near-field applications \cite{wangNearFieldIntegratedSensing2023}, \cite{galappaththigeNearFieldISACBeamforming2024}, \cite{sunBeamFocusingNearFieldISAC2024}. 
In \cite{wangNearFieldIntegratedSensing2023}, the authors consider a near-field ISAC scenario and utilize multiple signal classification (MUSIC) algorithm to maximize the sensing, while ensuring minimum communication rate. In \cite{galappaththigeNearFieldISACBeamforming2024}, the authors study multi-target detection in near-field ISAC system. Specifically, optimal transmit and receive beamforming is designed to minimize the transmit power of the base station (BS) while satisfying communication and sensing rate targets. The authors in \cite{sunBeamFocusingNearFieldISAC2024} study the design of near-field beam-focusing to maximize the minimum gain in beam pattern for radar detection while ensuring communication requirements are met. 

In the aforementioned works \cite{wangNearFieldIntegratedSensing2023}, \cite{galappaththigeNearFieldISACBeamforming2024}, \cite{sunBeamFocusingNearFieldISAC2024}, the radar operates in a full-duplex (FD) manner \cite{9724187}, i.e., simultaneously receiving while transmitting radar signals. However, with FD radar, communication functionality is implemented only in the downlink, i.e., in a half-duplex (HD) manner \cite{heFullDuplexCommunicationISAC2023}. This limitation constrains spectral efficiency and leaves the potential of FD communication underexplored. To address this gap, this paper investigates a novel FD communication-enabled near-field ISAC framework.  In this setup, BS functions as both a radar and communication transceiver, enabling simultaneous uplink and downlink communication alongside radar operations. While this configuration enhances spectral efficiency, it introduces significant challenges, such as interference between sensing and communication functionalities and coupling between uplink and downlink transmissions. These complexities necessitate advanced interference management and optimization strategies to fully realize the benefits of FD communication in near-field ISAC scenarios.


Motivated by the aforementioned discussion, we study FD communication in near-field ISAC systems. The objective is to evaluate the effectiveness and potential benefits of employing FD techniques in near-field scenarios, where communication and sensing functionalities are jointly integrated. Specifically, we develop a joint transmit/receive beamforming framework to minimize BS transmit power, while supporting simultaneous multi-user downlink, multi-user uplink, and multi-target sensing. The main contributions of this paper are summarized as follows:
\begin{itemize}
    \item To the best of the authors' knowledge, this is the first work to introduce FD communication in a near-field ISAC scenario. We formulate a joint transmit and receive beamforming problem to minimize the BS's transmit power while ensuring communication performance for both uplink and downlink users and achieving sensing performance for multiple targets.
    
    \item We derive the closed-form expressions for optimal receive beamformers to maximize the signal-to-interference-plus-noise ratio (SINR) for uplink communication and the SINR for targets detection, respectively.
    
    \item To address the non-convexity of the formulated beamforming problem, we employ an alternating optimization (AO) approach combined with semi-definite relaxation (SDR) and Rayleigh quotient techniques, enabling us to obtain a globally optimal solution.
    
    \item Numerical results are presented to validate the effectiveness of the proposed solution. Specifically, our results demonstrate that the BS can detect targets at identical angles—a capability not achievable in far-field scenarios. Additionally, we analyze the performance trade-off between target sensing and communications in near-field ISAC systems and highlight the superiority of FD-enabled near-field ISAC systems.
\end{itemize}

\section{System Model}
We consider a narrowband FD near-field ISAC system operating in the mmWave frequency with $N_t$ transmit and $N_r$ receive antennas, $K$ single-antenna downlink users and $L$ single-antenna uplink users. The BS also senses $M$ targets using the ISAC downlink signal. We assume that the BS antenna arrays are Uniform Linear Arrays (ULA) with antenna spacing $d$. This results in antenna aperture of $D = (N_t-1)d$, (respectively, $D=(N_r-1)d$) with a Rayleigh distance of $d = 2 D^2/\lambda$. Here $\lambda$ is the wavelength of the signal. Traditionally, Rayleigh distance defines the boundary between far-field and near-field, and we assume that the users and targets are present inside the near-field boundary \cite{7942128}. Furthermore, we assume that all Channel State Information (CSI) is perfectly available at the BS \cite{wangNearFieldIntegratedSensing2023}.

\subsection{Channel Model}

We use a near-field channel model for the channel between the users and the BS where the ULA center of the BS is the origin and user $k$ is assumed to be present at a distance of $r_k$ from the BS with angle $\theta_k$. 
The coordinate of the $n$-th element of the ULA is given by $\mathbf{s}_n = [0, nd], \forall n \in \{\frac{-N_t+1}{2}, \ldots, \frac{N_t-1}{2}\}$. The $k$-th user coordinate is thus given by $\mathbf{r_k} = [r_k \cos{\theta_k},r_k \sin{\theta_k}]$. The distance between $n$-th element of ULA and user $k$ is given by $r_{nk} = |\mathbf{r}_k-\mathbf{s}_n| = \sqrt{r_k^2 +n^2d^2 - 2 r_k n d \sin{\theta_k}}.$
\begin{figure}
    \centering
    \includegraphics[width=0.9\linewidth]{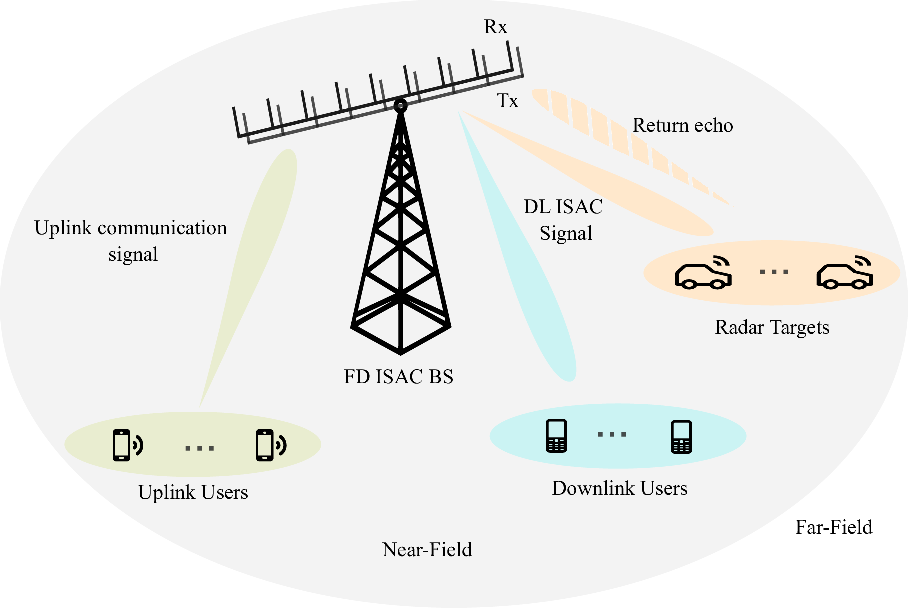}
    \caption{System Model for FD Near-Field ISAC: BS with $N_t = N_r =$ 65 antennas serving $K = $ 2 downlink users, $L = $ 2 uplink users, and sensing $M = 2$ targets.}
    \label{fig:system-model}
\end{figure}
Therefore, the channel between $n$-th element of ULA and user $k$ is given by \cite{10220205}
\begin{equation}
    h_{nk} = \sqrt{\beta_{nk}} e^{-j \frac{2\pi}{\lambda} r_{nk}},
\end{equation}
where $\beta_{nk}$ represents the free space path loss. The near-field channel vector $\mathbf{h}_k\in \mathcal{C}^{N_t \times 1}$ between the BS and the user is given by
\begin{equation}
    \mathbf{h}_k = \left[h_{\frac{-N_t+1}{2} k}, \ldots, h_{\frac{N_t-1}{2} k}\right]^T = \sqrt{\beta_k} \mathbf{a}(r_k, \theta_k),
\end{equation}
where it is assumed that the channel gain of each link between antenna elements and the user or target is approximately identical \cite{bjornsonPrimerNearFieldBeamforming2021} and $\mathbf{a}(r_k,\theta_k)$ is the near-field array response vector whose $n$-th element is given by $[\mathbf{a}(r_k,\theta_k)]_n = e^{-j \frac{2\pi}{\lambda} r_{nk}}$.

\subsection{Signal Model}
We consider a coherent time block \(T\), during which both communication and sensing parameters remain constant. The downlink signal \(\mathbf{x} \in \mathbb{C}^{N_t \times 1}\) for simultaneous downlink transmission and target sensing is expressed as \cite{9124713}:
\begin{equation}
    \mathbf{x} = \sum_{k \in \mathcal{K}} \mathbf{f}_k c_k + \sum_{m \in \mathcal{M}} \mathbf{s}_m,
\end{equation}
where \(\mathbf{f}_k \in \mathbb{C}^{N_t \times 1}\) represents the digital beamformer for the \(k\)-th downlink user, \(c_k \in \mathbb{C}\) denotes the transmission symbol for the \(k\)-th downlink user with unit power, i.e., \(\mathbb{E}[|c_k|^2] = 1\), and \(\mathbf{s}_m \in \mathbb{C}^{N_t\times1}\) represents the sensing signal for the \(m\)-th target. The covariance matrix of the sensing signal is given by \(\mathbf{S}_m \triangleq \mathbb{E}[\mathbf{s}_m\mathbf{s}_m^H]\). The signals \(c_k\) and \(\mathbf{s}_m\) are assumed to be independent of each other. The transmit beamforming design involves optimizing the sets of beamformers, i.e., \(\{\mathbf{f}_k\}_{k=1}^{K}\) and \(\{\mathbf{s}_m\}_{m=1}^{M}\).
The covariance matrix of the transmit signal is defined as:
\begin{equation}
    \mathbf{R}_x \triangleq \mathbb{E}\{\mathbf{x} \mathbf{x}^{H}\} = \sum_{k\in \mathcal {K}} \mathbf{f}_k \mathbf{f}_k^H + \sum_{m\in \mathcal {M}} \mathbf{S}_m.
\end{equation}

The communication signal received at the downlink user $k$ is given by
\begin{equation}    \label{eq:y_k}
\begin{split}
    {y}_{k}^\text{UE} &= \mathbf{h}_k^H \mathbf{x} + \mathbf{z}_k,\\
    {y}_{k}^\text{UE} &= \mathbf{h}_k^H \mathbf{f}_k c_k + \sum_{i=1, i \neq k}^{K}\mathbf{h}_k^H \mathbf{f}_i c_t + 
    \sum_{m=1}^{M} \mathbf{h}_k^H \mathbf{s}_m + {z}_k,
\end{split}
\end{equation}
where $\mathbf{h}_k\in\mathcal{C}^{N_t\times1}$ is the channel between the BS and downlink user $k$ and ${z}_k$ denotes the additive noise at the downlink user $k$, with zero mean and variance of $\sigma^2_k$.

Meanwhile, BS receives reflections from \(M\) targets and communication signals from \(L\) uplink users. Let \(c_l\) denote the signal transmitted by the \(l\)-th uplink user, where \(l \in \{1, \ldots, L\}\), with an average transmit power \(\mathbb{E}\{|c_l|^2\} = p_l\). The uplink channel between the BS and the \(l\)-th user is represented by \(\mathbf{h}_l \in \mathbb{C}^{N_r \times 1}\). Therefore, the multi-user uplink signal received at the BS can be expressed as $\sum_{l=1}^L \mathbf{h}_l c_l $.


To model the target echo signals, we assume a line-of-sight (LoS) path between $M$ targets and the BS. Let $\mathbf{G}_m \in \mathbb{C}^{N_r \times N_t}$ denote the near-field round-trip channel matrix for the $m$-th target, expressed as $\mathbf{G}_m \triangleq \sqrt{\zeta_m} \mathbf{a}_r(r_m, \theta_m) \mathbf{a}_t^H(r_m, \theta_m)$, where $\zeta_m$ represents the path loss and reflection power coefficient of the $m$-th target. The vectors $\mathbf{a}_t(r_m, \theta_m)$ and $\mathbf{a}_r(r_m, \theta_m)$ denote the transmit and receive array response vectors, respectively. Here, $r_m$ and $\theta_m$ represent the distance and angle of the $m$-th target relative to the BS. It is assumed that $\zeta_m$, $r_m$, and $\theta_m$ are either known or can be accurately estimated at the BS \cite{heFullDuplexCommunicationISAC2023}.

The composite signal received at the BS is expressed as:
\begin{equation} \label{eq:y_BS}
    \mathbf{y}^\text{BS} = \sum_{l=1}^L \mathbf{h}_l c_l + \sum_{m=1}^M \mathbf{G}_m \mathbf{x} + \mathbf{G}_\text{SI} \mathbf{x} + \mathbf{z}^\text{BS},
\end{equation}
where \(\mathbf{G}_{\text{SI}} \in \mathbb{C}^{N_r \times N_t}\) represents the self-interference (SI) channel caused by FD operations. While self-interference can be mitigated using successive interference cancellation (SIC) techniques in ISAC systems, it cannot be completely eliminated due to the limited dynamic range of the receiver \cite{9363029}. The term \(\mathbf{z}^\text{BS}\) denotes additive noise at the BS, assumed to have zero mean and variance of \(\sigma^2_\text{BS} \mathbf{I}_{N_r}\).

\subsection{Communication and Sensing SINR}
The performance of the ISAC system largely depends on radar and communication SINRs. Specifically, for point target detection in multiple-input multiple-output (MIMO) radars, detection probability is generally a monotonically increasing function of target SINR. Consequently from \eqref{eq:y_k}, the SINR for the downlink user $k$ is given by
\begin{equation}    \label{eq:SINR_DL}
    \gamma_k^{\text{DL}} = \frac{|\mathbf{h}_k^H \mathbf{f}_k|^2}{\sum_{i=1, i \neq k} ^K |\mathbf{h}_k^H \mathbf{f}_i|^2 + \sum_{m=1}^M \mathbf{h}_k^H \mathbf{S}_m \mathbf{h}_k + \sigma^2_k}.
\end{equation}

We apply the receive beamformers $\mathbf{w}_l \in \mathcal{C}^{N_r \times 1}$ on the received signal $\mathbf{y}^\text{BS}$ to decode uplink user data. Therefore, based on \eqref{eq:y_BS}, we obtain the uplink communication SINR for $l$-th user as

\begin{equation}   \label{eq:SINR_UL}
    \gamma_l^{\text{UL}} = \frac{p_l \mathbf{w}_l^H \mathbf{H}_l \mathbf{w}_l}   {\mathbf{w}_l^H\left(\sum_{i=1, i \neq l}^L p_i \mathbf{H}_i + \mathbf{A} \mathbf{R}_x \mathbf{A}^H + \sigma^2_{\text{BS}} \mathbf{I}_{N_r}\right)\mathbf{w}_l},
\end{equation}
where $\mathbf{A} \triangleq \sum_{m=1}^M \mathbf{G}_m+  \mathbf{G}_\text{SI}$ represents the interference due to the sensing signal and SI channel and $\mathbf{H}_i \triangleq \mathbf{h}_i \mathbf{h}_i^H$. Similarly, by applying another set of receive beamformers $\mathbf{u}_m \in \mathcal{C}^{N_r \times 1}$ to capture the desired target signals, the sensing SINR based on \eqref{eq:y_BS} is given by
\begin{equation}
    \Gamma_m  = \frac{\mathbf{u}_m^H \mathbf{G}_m \mathbf{R}_x \mathbf{G}_m^H \mathbf{u}_m}   {\mathbf{u}_m^H \left( \sum_{l=1}^L p_l \mathbf{H}_l  + \mathbf{B} \mathbf{R}_x \mathbf{B}^H + \sigma^2_{\text{BS}} \textbf{I}_{N_r}\right) \mathbf{u}_m},
\end{equation}
where $\mathbf{B} \triangleq \sum_{i=1, i\neq m }^M \mathbf{G}_i + \mathbf{G}_\text{SI}$ represents interference due to sensing signals from other targets and SI channel.

\section{Problem Formulation and Proposed Solution}
\subsection{Problem Formulation}
We aim to minimize the transmit power at the BS by jointly optimizing the transmit beamformers, \(\mathbf{f}_k\) and \(\mathbf{S}_m\), and the receive beamformers, \(\mathbf{w}_l\) and \(\mathbf{u}_m\), while ensuring that the minimal SINR requirements for uplink communications, downlink communications, and target sensing are satisfied. Let \(\mathcal{A} \triangleq \left\{ \{\mathbf{w}_l\}_{l=1}^L, \{\mathbf{u}_m\}_{m=1}^M, \{\mathbf{f}_k\}_{k=1}^K, \{\mathbf{S}_m\}_{m=1}^M \right\}\) represent the set of optimization variables. The corresponding optimization problem is formulated as:
\begin{subequations}
\begin{align}
\mathcal{P} =\; & \underset{\mathcal{A}}{\text{minimize}} 
&& \sum_{k=1}^{K} \Vert\mathbf{f}_k \Vert^2 + \sum_{m=1}^{M} \mathrm{Tr}(\mathbf{S}_m), \label{eq:P1} \\
& \text{subject to} 
&& \gamma_k^{\text{DL}} \geq \tau_k^{\text{DL}}, \quad \forall k, \label{eq:10b} \\
&&& \gamma_l^{\text{UL}} \geq \tau_l^{\text{UL}}, \quad \forall l, \label{eq:10c} \\
&&& \Gamma_m \geq \tau_m, \quad \forall m. \label{eq:10d}
\end{align}
\end{subequations}

where \(\tau_k^{\text{DL}}\) and \(\tau_l^{\text{UL}}\) denote the minimum required SINR for the \(k\)-th downlink user and \(l\)-th uplink user, respectively, and \(\tau_m\) represents the minimum required SINR for successfully sensing the \(m\)-th target.

\subsection{Proposed Solution}

The optimization problem \(\mathcal{P}\) is non-convex due to the coupling of optimization variables and the complexity of the SINR constraints. To address this, we adopt an AO approach \cite{64886}. Specifically, we first solve for the receive beamformers while keeping the transmit beamformers fixed. Then, with the optimized receive beamformers held constant, we optimize the transmit beamformers. This process is repeated iteratively until convergence.

\subsubsection{Receive Beamformer Solution}

The objective function of \(\mathcal{P}\) does not depend directly on the receive beamformers \(\mathbf{w}_l\) and \(\mathbf{u}_m\). Moreover, with fixed transmit beamformers \(\mathbf{f}_k\) and \(\mathbf{S}_m\), the receive beamformers \(\mathbf{w}_l\) and \(\mathbf{u}_m\) only influence the uplink SINR \(\gamma_l^{\text{UL}}\) and sensing SINR \(\Gamma_m\), respectively. Hence, to minimize the resultant transmit power, we maximize the corresponding SINRs. The subproblem for optimizing the receive beamformers is formulated as:

\begin{equation}
\max_{\mathbf{w}_l}~ \gamma_l^{\text{UL}},\quad \forall l,\label{eq:P2}
\end{equation}
\begin{equation}
\max_{\mathbf{u}_m}~ \Gamma_m,\quad \forall m.\label{eq:P3}
\end{equation}

\textit{Proposition 1:} The optimal receive beamforming solutions are given as follows:
\begin{align}
    \mathbf{w}_l^* &= \!\left( \sum_{i=1, i \neq l}^L \!\!\! p_i \mathbf{h}_i\mathbf{h}_i^H \!+\! \mathbf{A} \mathbf{R}_x \mathbf{A}^H \!+\! \sigma^2_{\text{BS}} \mathbf{I}_{N_r} \!\right)^{-1}\!\!\! \mathbf{h}_l, \quad\forall l, \label{eq:w_l} \\
    \mathbf{u}_m^* &= \!\left(  \sum_{l=1}^L \! p_l \mathbf{h}_l \mathbf{h}_l^H \!+\! \mathbf{B} \mathbf{R}_x \mathbf{B}^H  \!+\! \sigma^2_{\text{BS}} \mathbf{I}_{N_r} \!\right)^{-1}\!\!\! \Tilde{\mathbf{g}}_m, \quad\forall m, \label{eq:u_m}
\end{align}
where $\Tilde{\mathbf{g}}_m =\zeta_m \mathbf{a}_{r}(s_m,\theta_m)$.

\textit{Proof:} Based on the SINR expression for uplink communication in (\ref{eq:SINR_UL}), the receive beamforming problem in (\ref{eq:P2}) is a generalized Rayleigh quotient problem. By invoking the results from \cite{golubMatrixComputations2013}, the optimal solution for $\mathbf{w}_l$ is directly obtained as \eqref{eq:w_l}. 

For the sensing problem in (\ref{eq:P3}), note that the transmit term $\mathbf{a}_{t}^H(s_m,\theta_m) \mathbf{R}_x \mathbf{a}_{t}(s_m,\theta_m)$ is non-negative and independent of the receive beamformer $\mathbf{u}_m$. Thus, it can be ignored during optimization. We apply the generalized Rayleigh quotient result to $
\frac{
    \mathbf{u}_m^H \zeta_m\mathbf{a}_{r}(s_m,\theta_m) \mathbf{a}_{r}^{H}(s_m,\theta_m) \mathbf{u}_m
}{
    \mathbf{u}_m^H \left(
        \sum_{l=1}^L p_l \mathbf{h}_l \mathbf{h}_l^H
        + \mathbf{B} \mathbf{R}_x \mathbf{B}^H
        + \sigma_{\text{BS}}^{2} \mathbf{I}_{N_r}
    \right) \mathbf{u}_m
}
$
to arrive at the optimal solution for $\mathbf{u}_m$ in (\ref{eq:u_m}), which is a minimal mean-squared error (MMSE) filter.

\subsubsection{Transmit Beamformer Solution}
With the optimal receive beamformers (\(\mathbf{w}_l^*\) and \(\mathbf{u}_m^*\)) fixed, we optimize the transmit beamformers using SDR. We define auxiliary variables as $\mathbf{F}_k \stackrel{\triangle}= \mathbf{f}_k \mathbf{f}_k^H$, where $\mathbf{F}_k$ is a positive semidefinite matrix with a rank one constraint. Relaxing by dropping the non-convex rank-one constraints, the problem becomes: 
\begin{subequations} \label{eq:P_full}
\begin{align}
    & \qquad  \quad\min_{\substack{\mathbf{F}_k \succeq 0 \\ \mathbf{S}_m \succeq 0}}
    \quad \sum_{k=1}^K \mathrm{Tr}(\mathbf{F}_k) + \sum_{m=1}^M \mathrm{Tr}(\mathbf{S}_m),
      \\
    & \qquad \text{s.t.}\, 
    \sum_{i=1,\, i\neq k}^K \mathrm{Tr}(\mathbf{H}_k \mathbf{F}_i)
    + \sum_{m=1}^M \mathrm{Tr}(\mathbf{H}_k \mathbf{S}_m)
    + \sigma_k^2 \nonumber \\
    &\qquad  \qquad  \qquad 
    - \frac{\mathrm{Tr}(\mathbf{H}_k \mathbf{F}_k)}{\tau_k^{\mathrm{DL}}}
    \leq 0,\quad \forall k, \label{Const1}\\
    &\qquad 
    \sum_{i=1,\, i\neq l}^L p_i\, \operatorname{Tr}(\mathbf{H}_i \mathbf{W}_l)
    + \operatorname{Tr}\left( \overline{\mathbf{R}}_x\, \mathbf{A}^H \mathbf{W}_l \mathbf{A} \right)
     \nonumber \\
    &\qquad 
    + \sigma_{\mathrm{BS}}^2\, \| \mathbf{w}_l\|^2 - \frac{p_l}{\tau_l^{\mathrm{UL}}} \operatorname{Tr}(\mathbf{H}_l \mathbf{W}_l)
    \leq 0, \quad \forall l, \label{Const2}\\
    &\qquad 
    \sum_{l=1}^L p_l\, \operatorname{Tr}(\mathbf{H}_l\, \mathbf{U}_m)
    + \operatorname{Tr}\left( \overline{\mathbf{R}}_x\, \mathbf{B}^H \mathbf{U}_m \mathbf{B} \right)+  \nonumber \\
    &\; \:
    \sigma_{\mathrm{BS}}^2\, \| \mathbf{u}_m\|^2
    - \frac{1}{\tau_m} \operatorname{Tr}\left( \overline{\mathbf{R}}_x\, \mathbf{G}_m^H \mathbf{U}_m \mathbf{G}_m \right )
    \leq 0, \quad \forall m, \label{Const3}
\end{align}
\end{subequations}
where $\mathbf{\overline{R}}_x \triangleq  \sum_{k\in \mathcal {K}}  \mathbf{F}_k +  \sum_{m\in \mathcal {M}} \mathbf{S}_m$, $\mathbf{W}_l \triangleq \mathbf{w}_l \mathbf{w}_l^H$ and $\mathbf{U}_m \triangleq \mathbf{u}_m \mathbf{u}_m^H$, respectively, while also using the cyclic permutation property of the trace.  

The relaxed problem is a semidefinite programming (SDP) problem, which can be solved globally using methods such as the interior point method or off-the-shelf convex optimization tools like CVX.


Let the globally optimal solutions obtained by solving the relaxed problem (\ref{eq:P_full}) be denoted as \(\mathbf{F}_k^*\) and \(\mathbf{S}_m^*\). If these solution matrices have a rank of one, then the SDR solution is also optimal for the original problem (\ref{eq:P1}), and relaxing the rank-one constraint does not affect global optimality. However, if the solution matrices are not rank-one, we apply the following result from \cite[Theorem 1]{9124713}.

\textit{Theorem 1:} There exist rank-one solutions to the relaxed problem, denoted as \(\mathbf{F}_k^{**}\) and \(\mathbf{S}_m^{**}\), which are constructed from the solutions \(\mathbf{F}_k^*\) and \(\mathbf{S}_m^*\) while achieving the same performance bounds as the relaxed rank-one constraint. These solutions can be constructed as follows:
\begin{align}
    \mathbf{f}_k^{**} &= (\mathbf{h}_k^H \mathbf{F}_k^* \mathbf{h}_k)^{-\frac{1}{2}} \mathbf{F}_k^* \mathbf{h}_k, \nonumber \\
    \mathbf{F}_k^{**} &= \mathbf{f}_k^{**} (\mathbf{f}_k^{**})^H, \nonumber \\
    \sum_m \mathbf{S}_m^{**} &= \mathbf{R}_x^{*} - \sum_k \mathbf{f}_k^{**} (\mathbf{f}_k^{**})^H, \label{eq:FK}
\end{align}
where the optimal beamforming vectors for sensing is given by:
\(
    \mathbf{s}_m = \sqrt{\lambda_m} \mathbf{v}_m,
\)
where \(\lambda_m\) and \(\mathbf{v}_m\) denote the eigenvalue and corresponding eigenvector, respectively.

\textit{Proof:} Please refer to Appendix for details.




The procedure to compute the beamforming vectors is summarized in Algorithm~1.

\begin{algorithm}
\caption{AO-Based Beamforming Algorithm}
\begin{algorithmic}[1]  
    \State \textbf{Initialization:} Initialize a feasible solution set 
    $\left\{ \{ \mathbf{f}_k \}_{k=1}^{K}, \{ \mathbf{s}_m \}_{m=1}^{M} \right\}$, 
    iteration index $i = 0$, and convergence accuracy $\epsilon$
    
    \State \textbf{repeat}
    
    \State Increment iteration index: $i = i + 1$
    
    \State Solve for the receive beamformers 
    $\mathbf{w}_l^{(i+1)}$ and $\mathbf{u}_m^{(i+1)}$ using equations~(13) and (14)
    
    \State Compute the optimal transmit solutions 
    $\mathbf{F}_k$ and $\mathbf{S}_m$ by solving the relaxed problem in equation~(15)
    
    \State \textbf{until Convergence}
    
    \State Calculate rank-one solutions for transmit beamformers using equation~(16)
    
    \State \textbf{Output:} Optimal beamforming solutions: $\mathcal{A}$
\end{algorithmic}
\end{algorithm}

\section{Simulation Results}
We evaluate the proposed algorithm via simulations with uniform linear arrays (ULAs) at both transmitter and receiver, each with $N_t = N_r = 65$ antennas at 28 GHz. We use a free-space path-loss model for the large-scale fading. The BS serves $K = 2$ downlink users, $L = 2$ uplink users, and senses $M = 2$ targets, all in its near-field region. The required SINR thresholds for target sensing, uplink communication and downlink communications are set to $\tau_m = 20 \text{ dB}, \forall m, \tau_l^{\text{UL}} = 12 \text{ dB}, \forall l, \tau_k^{\text{DL}} = 12 \text{ dB}, \forall k$, respectively. Noise power at the BS and each downlink user is set to $-94 \text{ dBm}$, while the transmit power of each uplink user is fixed at -30 dBm. The target, downlink, and uplink directions are set to $\{0^\circ, 30^\circ\}$, $\{-40^\circ, 60^\circ\}$, and $\{45^\circ, -65^\circ\}$, respectively, with all entities positioned at a uniform radial distance of 9–10 meters from the base station.

We compare three baselines:

\subsubsection{HD ISAC} A HD time-division duplex (TDD) scheme where downlink and uplink communications occur in separate time slots while sensing is performed continuously at the BS. In the downlink slot, the BS transmits an ISAC signal $\mathbf{x}$ containing both downlink user and target signals  while receiving radar returns through receive beamformer $\mathbf{u}_m$, In the uplink slot, the BS transmits only the radar signal while simultaneously receiving both radar and uplink user signals. SINR thresholds are adjusted for rate fairness, i.e., $\tau_i^{\text{HD}} = (1+ \tau_i^{\text{FD}})^2 - 1$ and power is averaged over both slots.

\subsubsection{Communication-only} Radar SINR constraints are omitted to evaluate communication performance alone.

\subsubsection{Far-field (FF) ISAC} This benchmark scheme allows us to quantify the degradation in system performance when a far-field channel model is incorrectly applied in near-field ISAC scenarios. The channel response follows $h_{nk} = \sqrt{\beta_{nk}} e^{j \frac{2\pi}{\lambda} n d \sin{\theta}}.$ 

First, we show the beampattern gain achieved by our proposed algorithm. The optimized radar downlink signal $\mathbf{x}^*$ and receive $\mathbf{u}_m^*$ beamformers jointly produce a combined beampattern gain \cite{heFullDuplexCommunicationISAC2023} defined as 
\begin{equation}
    p_m(r_m, \theta_m) = | (\mathbf{u}_m^*)^H \mathbf{a}_r(r_m, \theta_m) \mathbf{a}_t^H(r_m, \theta_m)  \mathbf{x}^*|^2. 
\end{equation}

\begin{figure}
    \centering
    \includegraphics[width=0.85\linewidth]{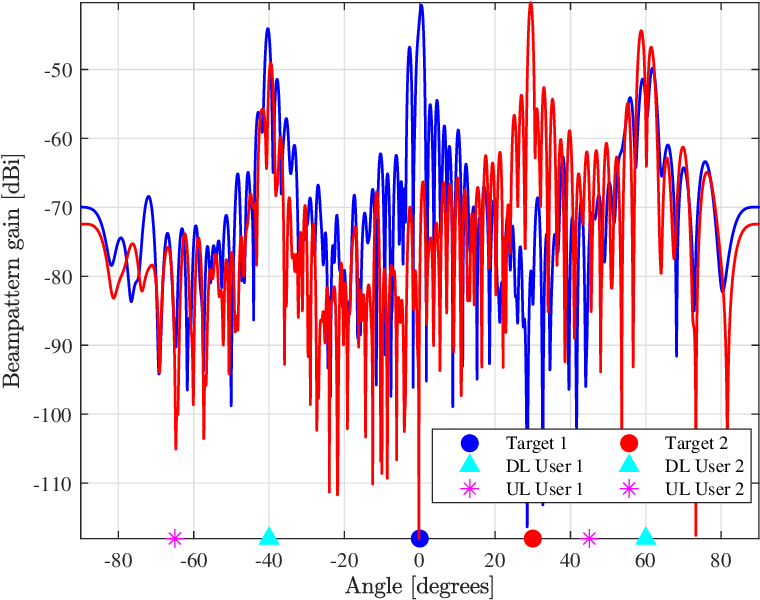}
    \caption{Combined Beampattern for Targets and Downlink Users: Beams towards targets at $\{0^\circ, 30^\circ\}$ and downlink users at $\{-40^\circ, 60^\circ\}$, with nulls at uplink user directions.}
    \label{fig:beam_rad_all}
\end{figure}

Fig. \ref{fig:beam_rad_all} shows the combined beampattern gain achieved by the transmit and receive beamformers. The beams are pointed towards the two targets and two downlink users, respectively, while placing nulls in the directions of uplink users to mitigate interference with radar sensing. Near-field beam focusing reduces the need for deep nulls compared to the far-field scenario due to less interference. 
\begin{figure}
    \centering
    \includegraphics[width=0.85\linewidth]{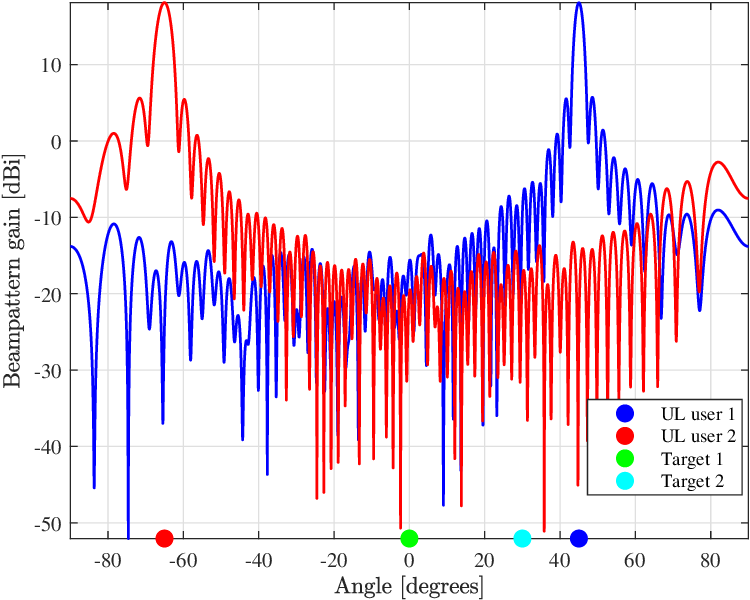}
    \caption{Uplink receive beampattern: Each beamformer steers its main lobe to its uplink user and nulls interference from other users and targets.}
    \label{fig:Beam_UL}
\end{figure}

The receive beampattern for uplink user $l$ based on the receive beamformer $\mathbf{w}_l^*$ is defined as $p_l(r_l,\theta) = |(\mathbf{w}_l^*)^H \mathbf{a}_r(r_l,\theta_l)|^2$. Fig. \ref{fig:Beam_UL} illustrates that each beamformer places a main lobe toward its corresponding uplink user while nullifying interference from other users and targets. This demonstrates the effectiveness of the proposed design in isolating uplink signals. 

\begin{figure}
    \centering
    \includegraphics[width=0.85\linewidth]{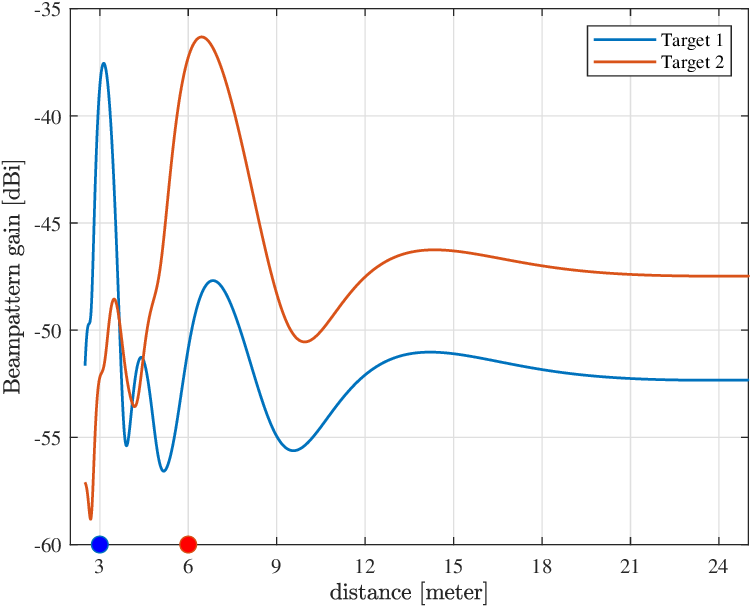}
    \caption{Near-field ISAC beampattern gain at $0^\circ$: Detects multiple targets at the same angle but at 3 m and 6 m.}
    \label{fig:Beam_dist}
\end{figure}

To highlight near-field beam-focusing capability in detecting targets at the same angle but different distances, we place two targets at $0^\circ$ but at 3 m and 6 m. Fig. \ref{fig:Beam_dist} shows two distinct peaks corresponding to these distances, confirming that near-field ISAC can effectively detect multiple targets in the same direction by leveraging distance-dependent beam focusing.

\begin{figure}
    \centering
    \includegraphics[width=0.85\linewidth]{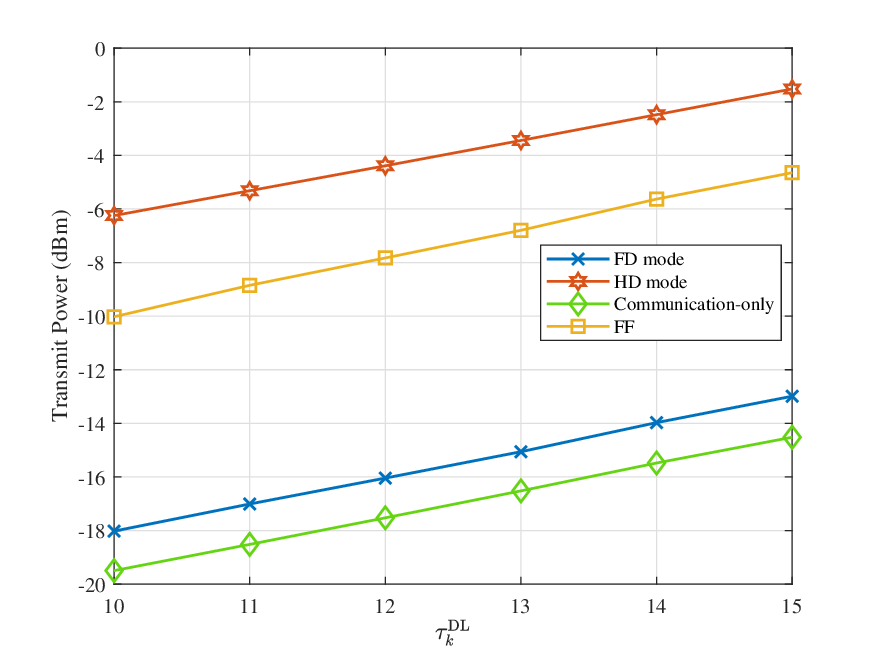}
    \caption{Transmit power vs. downlink SINR threshold for FD, HD, communication-only, and far-field ISAC scenarios.}
    \label{fig:Comparison}
\end{figure}

Fig. \ref{fig:Comparison} illustrates the optimal transmit power versus downlink user SINR thresholds. The power consumption across FD, HD, communication-only, and FF modes exhibits an increasing trend with higher downlink user SINR requirements. The FD mode demonstrates superior performance by consuming approximately 12 dB less power compared to the HD mode at equivalent SINR levels. 
 The far-field modelling assumption requires approximately 8 dB higher transmit power due to increased interference and distributed beam energy. 
This significant power reduction in the near-field scenario validates the effectiveness of our proposed beamforming design in satisfying the SINR constraints.



\section{Conclusion}

This paper presents a novel framework for FD near-field ISAC systems, that demonstrates significant advantages in power efficiency and target detection capabilities. By jointly optimizing transmit and receive beamformers, our proposed algorithm minimizes transmit power while ensuring reliable target sensing and communication performance. Numerical results validate the effectiveness of our approach, 
specifically, the power consumption in FD mode is reduced by approximately 12 dB compared to HD schemes, demonstrating the superiority of FD communication in near-field ISAC systems. Furthermore, our results demonstrate the unique capability of near-field ISAC to detect multiple targets at identical angles through beam-focusing. 

\appendix


\subsection*{A. Proof of \textit{Theorem 1}}
Extending the proof from \cite[Theorem 1]{9124713}, we demonstrate that \(\mathbf{F}_k^{**}\) and \(\mathbf{S}_m^{**}\) constitute a feasible solution to the original problem (\ref{eq:P1}) while achieving equivalent performance to the rank-one relaxed solution \(\mathbf{F}_k^*\) and \(\mathbf{S}_m^*\).

First, note that \(\mathbf{F}_k^{**}\) is positive semidefinite and rank-one by construction. We prove that \(\mathbf{F}_k^* - \mathbf{f}_k^{**} (\mathbf{f}_k^{**})^H \succeq 0\). For any arbitrary vector \(\mathbf{g} \in \mathbb{C}^{N_t \times 1}\):
\begin{equation} \label{eq:FK_1}
    \begin{split}
        \mathbf{g}^H (\mathbf{F}_k^* - \mathbf{f}_k^{**} ) \mathbf{g} &= \mathbf{g}^H \mathbf{F}_k^* \mathbf{g} - (\mathbf{h}_k^H \mathbf{F}_k^* \mathbf{h}_k)^{-1} |\mathbf{g}^H \mathbf{F}_k^* \mathbf{h}_k|^2 \\
        &\geq \mathbf{g}^H \mathbf{F}_k^* \mathbf{g} - \mathbf{g}^H \mathbf{F}_k^* \mathbf{g} = 0,
    \end{split}
\end{equation}
where the inequality follows from the Cauchy-Schwarz inequality: \(|\mathbf{g}^H \mathbf{F}_k^* \mathbf{h}_k|^2 \leq (\mathbf{g}^H \mathbf{F}_k^* \mathbf{g})(\mathbf{h}_k^H \mathbf{F}_k^* \mathbf{h}_k)\). Therefore, \(\mathbf{F}_k^* - \mathbf{f}_k^{**} (\mathbf{f}_k^{**})^H \succeq 0\).

Using this result and the fact that \(\sum_m \mathbf{S}_m^* \succeq 0\), it follows:
\begin{equation}
    \sum_m \mathbf{S}_m^{**} = \sum_m \mathbf{S}_m^* + \sum_k (\mathbf{F}_k^* - \mathbf{f}_k^{**} \mathbf{f}_k^{**H}) \succeq 0.
\end{equation}

Finally, we demonstrate:
\begin{equation} \label{eq:FK_2}
    \begin{split}
        \mathbf{h}_k^H \mathbf{F}_k^{**} \mathbf{h}_k &= \mathbf{h}_k^H \mathbf{f}_k^{**} \mathbf{f}_k^{**H} \mathbf{h}_k \\
        &= \mathbf{h}_k^H (\mathbf{h}_k^H \mathbf{F}_k^* \mathbf{h}_k)^{-1} \mathbf{F}_k^* \mathbf{h}_k \mathbf{h}_k^H \mathbf{F}_k^{*H} \mathbf{h}_k \\
        &= \mathbf{h}_k^H \mathbf{F}_k^* \mathbf{h}_k
    \end{split}
\end{equation}

From \eqref{eq:FK_2} and the fact that \(\mathbf{R}_x^* = \mathbf{R}_x^{**}\) from \eqref{eq:FK}, the constraints \eqref{Const1}, \eqref{Const2}, and \eqref{Const3} are satisfied. Furthermore, using the fact that $\sum_{k=1}^K \mathrm{Tr} \left( \mathbf{F}_k^{**}\right) + \sum_{m=1}^M  \mathrm {Tr} \left( \mathbf{S}_m^{**}\right) = \sum_{k=1}^K \mathrm{Tr} \left( \mathbf{F}_k^*\right) + \sum_{m=1}^M  \mathrm {Tr} \left( \mathbf{S}_m^*\right),$ the solution provided by \(\mathbf{F}_k^{**}\) and \(\mathbf{S}_m^{**}\) is also optimal for \eqref{eq:P_full}.
To construct the sensing vectors \(\mathbf{s}_m^{**}\), we perform eigendecomposition of \(\mathbf{S}_m^{**}\), such that:
$
\sum_m \mathbf{S}_m^{**} = \mathbf{V} \Lambda \mathbf{V}^H = \sum_m \lambda_m \mathbf{v}_m \mathbf{v}_m^H,
$
where the \(m\)-th largest eigenvector is selected as the sensing beamformer, expressed as:
$
\mathbf{s}_m = \sqrt{\lambda_m} \mathbf{v}_m.
$

\bibliographystyle{IEEEtran}
\bibliography{MyLibrary,NF}

\end{document}